\input{aipcheck}

\documentclass[
    ,final            
  ]
  {aipproc}

\usepackage{amsmath}
\usepackage{graphicx}
\def\lsim{\raise0.3ex\hbox{$\;<$\kern-0.75em\raise-1.1ex\hbox{$\sim\;$}}}
\def\gsim{\raise0.3ex\hbox{$\;>$\kern-0.75em\raise-1.1ex\hbox{$\sim\;$}}}

\layoutstyle{6x9}

\begin{document}

\title{$\gamma-$ray line generated by the Green-Schwarz mechanism}

\classification{95.30.Cq,95.35.+d,98.35.Gi,98.62.Gq}
\keywords      {Dark Matter, Detection, String Models}

\author{Y. Mambrini}{
  address={Laboratoire de Physique Th\'eorique,
Universit\'e Paris-Sud, F-91405 Orsay, France}
}

\begin{abstract}
 We study the phenomenology of a $U_X(1)$ extension of the Standard Model where the SM particles
are not charged under the new abelian group.
The Green-Schwarz mechanism insures that the model is anomaly free.
The erstwhile invisible dark gauge field $X$, even if
 produced with difficulty at the LHC has however a clear signature in gamma-ray telescopes.
 We investigate what BSM scale (which can be interpreted as a low-energy string scale)
  would be reachable by the FERMI/GLAST telescope after
 5 years of running and show that it can be highly competitive with the LHC.
\end{abstract}

\maketitle


\section{Introduction}

Recent experiments have shown that dark matter (DM) makes up about $25 \%$ of
the Universe's energy budget, but
its nature is not yet understood.
A natural candidate is a class of weakly coupled massive particle (WIMP). Such a
100 GeV candidate would naturally give the right order of magnitude for the thermal
relic abundance.
One of the most studied extensions of the Standard Model (SM) is the Minimal Supersymmetric Model MSSM
(mSUGRA in its local version) which extends the matter spectrum
thanks to an extension of the space-time
symmetries.
Alternative ideas have included attempts to study the dark matter consequences of
the simplest gauge extension of the SM by adding a hidden sector charged under a
new $U_X(1)$ symmetry  and some
specific signatures in gamma ray telescope were discussed in \cite{UprimeDMpheno,Arkani,Feldman:2007wj}.
Many of these models have exotic
matter, including non-vectorlike matter that couples to both SM and hidden sector gauge group,
the lighter of which would be the DM candidate.
However, none of these works focussed on the consequences of the anomalies induced by these kind of
spectra on DM detection. Recently, some works have studied extensively the LHC prospect of
such a construction \cite{KumarWells,Wellsteam} whereas one dimension 6 effective operator approach
leads to specific astrophysical signals \cite{Us}. Even more recently, the author of \cite{Me} shows
that low energy scale would give similar smoking gun signal at the FERMI telescope.
In this proceeding, we show that, even if the SM particles are not charged under the extra $U_X(1)$, anomalies
generated by the heavy fermionic spectrum can generate through the Green-Schwarz mechanism,
$XZ\gamma$ effective couplings. This coupling induces a clear $\gamma$ ray line signal observable by FERMI
at energy

\begin{equation}
E_\gamma= m_{DM} \left( 1- \frac{M_Z^2}{4 m_{DM}^2} \right),
\end{equation}

\noindent
where $m_{DM}$ is the WIMP mass. Line emission provides a feature that helps to discriminate against the background.



\section{Effective description of the model}

It is well known that any extension of the SM which introduces chiral fermions
with respect to gauge fields suffers from anomalies, a phenomenon of breaking of gauge
symmetries of the classical theory at one-loop level. Anomalies are responsible for instance for
a violation of unitarity and make a theory inconsistent \cite{ABJ,Coriano}.
For this reason if any construction introduces a new fermionic sector to address
the DM issue of the SM, it is vital to check the cancelation of anomalies
and its consequences on the Lagrangian and couplings.
In this letter, we concentrate on the Green-Schwarz mechanism  which arises automatically
in string theory settings. The idea is to add to the Lagrangian local gauge non-invariant
terms in the effective action whose gauge variations cancel the anomalous triangle diagrams.
There exist two kinds of term which can cancel the mixed $U_X(1)\times G_A^{\mathrm{SM}}$
anomalies, with $U_X(1)$ being the hidden sector gauge group and $G_A^{\mathrm{SM}}$ one
of the SM gauge group $SU(3)\times U(2)\times U_Y(1)$ :
the Chern Simons (CS) term which couples the $G_A^{\mathrm{SM}}$ to the $U_X(1)$ gauge boson,
and the Peccei-Quinn (PQ, or Wess-Zumino (WZ)) term which couples the $G_A^{\mathrm{SM}}$
gauge boson to an axion. In the effective action, these terms are sometimes called
Generalized Chern--Simons (GCS) terms \cite{Abdk}.
In order to describe the relevant structure,
we can separate the effective Lagrangian into a sum of classically gauge variant and gauge invariant
terms\footnote{We will consider $U_X(1) \times U_Y^2(1)$ mixed anomalies throughout the
paper to simplify the formulae, the generalization to $U_X(1) \times SU^2(N)$ being straightforward.}
\cite{KumarWells,Wellsteam,Abdk} :

\begin{eqnarray}
{\cal L}_{inv} &&= - \frac{1}{4 g'^2} F^{Y \mu \nu} F^Y_{\mu \nu}
- \frac{1}{4 g_X^2} F^{X \mu \nu} F^X_{\mu \nu}
-\frac{1}{2} (\partial_\mu a_X - M_X X_\mu)^2
-i \overline{\psi} \gamma^\mu D_\mu \psi
\nonumber
\\
{\cal L}_{var} &&=
\frac{C}{24 \pi^2} a_X \epsilon^{\mu \nu \rho \sigma} F^Y_{\mu \nu} F^Y_{\rho \sigma}
+\frac{E}{24 \pi^2} \epsilon^{\mu \nu \rho \sigma} X_{\mu} Y_{\nu} F^Y_{\rho \sigma}
\label{Lagrangian}.
\end{eqnarray}

\noindent
The Stueckelberg axion $a_X$ ensures the gauge invariance of the effective Lagrangian and
$g_X$ and $F^X_{\mu \nu}=\partial_{\mu} X_\nu - \partial_\nu X_\mu$
are the gauge coupling and field strength of $U_X(1)$. The axion
has a shift transformation under $U_X(1)$
\begin{equation}
\delta X_{\mu} \ = \ \partial_{\mu} \alpha \quad , \quad \delta a_X
\ = \ \alpha \  M_X .
\end{equation}

\noindent
Notice that our dark matter candidate is expected to be chiral with respect
to the dark sector $U_X(1)$, with a mass of the order of $M_X$
because its mass should be generated by the spontaneous breaking of $U_X(1)$.
This differs considerably from the result obtained with a leptophylic dark sector \cite{Ko}
 which considered vector-like dark matter with a Dirac mass term. A hierarchy
 between $X$ and the lightest heavy fermion charged under $U_X(1)$
 (our natural dark matter candidate) would imply a hierarchy between hidden sector
 Yukawa and gauge couplings. The hidden fermionic sector being chiral,
 looking for the effects of mixed anomalous diagrams on DM phenomenology is not an ad-hoc assumption,
  but a necessity.

From Eq.(\ref{Lagrangian}), we can now express our effective vertices in terms of  finite integrals
\cite{Abdk, Me, KumarWells}:


\begin{eqnarray}
\Gamma_{\mu \nu \rho}^{XZZ}&=&
-2 i \frac{\sin \theta_W^2 g_X g'^2 \mathrm{Tr[Q_X Q_Y Q_Y]}}{8 \pi^2 M^2}
\left(
[\tilde I_1 ~ p_2^2+ \tilde I_2~p_1.p_2]\epsilon_{\mu \nu \rho \sigma} p_1^\sigma
-[\tilde I_2 ~p_1.p_2+ \tilde I_1~p_1^2]\epsilon_{\mu \nu \rho \sigma} p_2^\sigma
\right.
\nonumber
\\
&+&
\left.
[\tilde I_1 ~p_{2\nu} + \tilde I_2 ~p_{1\nu}]\epsilon_{\mu\rho\sigma\tau}p_2^\sigma p_1^\tau
-[ \tilde I_2 ~p_{1\rho} + \tilde I_1 ~p_{2 \rho}]\epsilon_{\mu \nu \sigma \tau}
p_2^\sigma p_1^\tau
\nonumber
\right.
\\
&+& \left.\frac{1}{p_3^2} [\tilde I_1 ~ p_2^2 + \tilde I_2 p_1^2 + (\tilde I_1 + \tilde I_2 )p_1.p_2]
p_{3\mu}\epsilon_{\nu\rho\sigma\tau} p_2^\sigma p_1^\tau
\right)
\label{XZZcoupling}
\end{eqnarray}

\begin{eqnarray}
\Gamma_{\mu \nu \rho}^{X Z \gamma}&=&
2(\cos \theta_W / \sin \theta_W) \Gamma_{\mu \nu \rho}^{XZZ}
\label{XZgcoupling}
\end{eqnarray}

\begin{equation}
\Gamma^{X \psi_{DM} \psi_{DM}} = \frac{g_X}{4}\gamma^\mu
\left(
[q_X^R + q_X^L] + [q_X^R-q_X^L]\gamma^5
\right)
\label{XDMDMcoupling}
\end{equation}

\noindent
where we defined $\tilde I_i= M^2 I_i$, $\tilde I_i$ being a dimensionless integral
and $M$ the $U_X(1)$ breaking scale (typically the masses of the hidden fermions
running in the loops).  This scale can be thought of as coming from effective derivative couplings
as was explicitly shown in \cite{Wellsteam,Us}.



\section{Results}

In our analysis we use diffuse-model simulated data from the centre annulus (r $\in [20^0, 35^0]$),
excluding the region within $15^0$ of the Galactic plane. It has recently been shown
that it is possible in this case to minimize the contribution
of the Galactic diffuse emission and could give a signal-to-noise ratio up to 12 times greater
than at the Galactic Center (GC). A very interesting feature of excluding the GC in our analysis
lies in the fact that our results are quite unsensitive to the different dark matter profile
(Einasto, NFW or Moore) as their contributions
differ largely within the parsec region around the GC.
In addition, we use the LAT line energy sensitivity for 5$\sigma$
detection. The $\gamma$-ray spectrum is calculated
 using an adapted version\footnote{The author wants to thank warmfuly G. Belanger and S. Pukhov for the precious
 help concerning the modification of the code.} of micrOMEGAs \cite{Micromegas}.

\begin{figure}
  \includegraphics[height=.3\textheight]{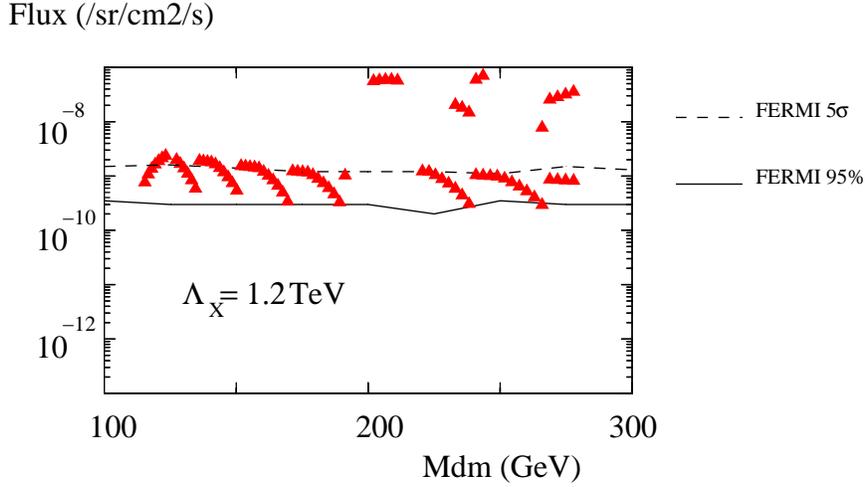}
  \caption{monochromatic $\gamma-$ray fluxes generated by Green-Scharz mechanism
in comparison with expected $5\sigma$ and $95\%$ CL sensitivity contours
(5 years of FERMI operation) for the conventional background and unknown WIMP energy,
for an effective scale $\Lambda_X=1.2$ TeV \cite{Me}}
\label{fig:spectrum}
\end{figure}

\begin{figure}
  \includegraphics[height=.3\textheight]{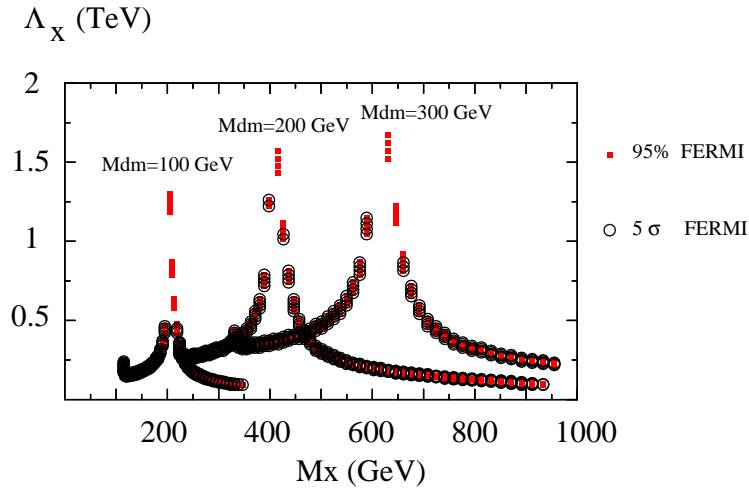}
  \caption{Monochromatic $\gamma-$ray fluxes generated by Green-Scharz mechanism
in comparison with expected $5\sigma$ and $95\%$ CL sensitivity contours
(5 years of FERMI operation) for the conventional background and unknown WIMP energy,
for different values of the scale $\Lambda_X$ \cite{Me}.}
\label{fig:Lbsm}
\end{figure}

We calculated the fluxes generated by the Green-Schwarz mechanism and compared it with the
expected sensitivity of FERMI after 5 years of data-taking.
The results are presented
in Fig.\ref{fig:spectrum}.
We clearly see
that for $\Lambda_X = 1.2$ TeV, all the parameter space would be observable
by FERMI at 95\% CL. Indeed, the points that respect the WMAP
constraints lie around the p\^ole $M_X \sim 2 M_{DM}$ where
$\sim 60$\% of the annihilation rate is dominated by the $Z\gamma$ final state.
This proportion still holds for annihilating DM in the Galactic
halo and gives a monochromatic line observable by FERMI. We also show what
values of $\Lambda_X$ are accessible by measurement of
$\gamma-$ray for different values of DM masses (100, 200 and 300 GeV)
in Fig. \ref{fig:Lbsm}, where all the points respect the
WMAP relic abundance. We can see that for $m_{DM}\gsim 100$ GeV,
$\Lambda_X \gsim 1$ TeV the model still gives a signal observable by FERMI at 95\% CL.
Obviously, for points lying away from the s-resonance
we still have points which can respect WMAP constraint for lower
values of $\Lambda_X$. All these points (black circles in
Fig. \ref{fig:Lbsm}) would be observable at 5$\sigma$ after 5 running years of
FERMI.

\section{Discussion}

Some models exhibit specific signatures as high energy rise due to final state radiation
\cite{Barger,Gammaradiation}, but need large positron excess.
Other dark matter candidates (like the Inert Higgs Dark Matter
(IHDM)
\cite{IHDMline} or extra-dimensional chiral square theories \cite{Chiralsquare})
can give strong monochromatic
signals from $\gamma \gamma$ or $Z \gamma$ final states.
However, except in \cite{Us, Me}, none of these models exhibit only one $\gamma$ ray line but two or three.
In the Green--Schwarz mechanism, $\gamma \gamma$ final state is excluded by spin conservation,
we are thus left with one monochromatic line, which would be a clear signature of the model.
One possibility would be a confusion in the signal between a decaying Dark Matter \cite{Decaying} and an
annihilating one \cite{Me}.

Concerning the colliders expectations,
The Large Hadron Collider phenomenology of couplings generated by anomalous extra $U_X(1)$
 has recently been studied in the framework of the Green-Schwarz mechanism
 \cite{KumarWells} and higher dimensional operators \cite{Wellsteam}.
The former computed the production cross-section
of the $X$ boson from vector boson fusion at the LHC. It was shown that,
for the mass range ($M_X \sim 500-1000$ GeV), LHC could
detect the new physics through $pp \rightarrow X \rightarrow ZZ \rightarrow 4l$
processes provided $\Lambda_X \sim 100-150$ GeV. In the case of $\gamma-$ray detection
we showed that in the same framework, the FERMI satellite will be much more efficient
and will be able to probe a scale $\Lambda_X \sim 100-2000$ GeV.
The main reason is that the production of the $X$ boson occurs through vector boson
fusion and the $qqX$ coupling is suppressed by a factor $\sim g_X/\Lambda_X^2$, whereas in the
case of DM annihilation, the $\psi_{DM} \psi_{DM}X$ coupling is directly proportional
to $g_X$.


\section*{Acknowledgments}{
The author want to thank particularly E. Dudas, S. Abel, P. Ko, A. Morselli
and A. Romagnoni for very useful discussions.  He is also grateful to
G. Belanger and S. Pukhov for substantial help with micrOMEGAs.
Likewise, the author
would like to thank T. Gherghetta for having him discovered the pleasure of Australian cooking,
and C. Blin for his constant support.
}

\end{document}